\documentstyle[aps,epsfig,multicol]{revtex}

\newcommand{\be}{\begin{equation}}
\newcommand{\ee}{\end{equation}}
\newcommand{\bea}{\begin{eqnarray}}
\newcommand{\eea}{\end{eqnarray}}
\newcommand{\lb}{\left[}
\newcommand{\rb}{\right]}
\newcommand{\lp}{\left(}
\newcommand{\rp}{\right)}

\begin{document}
\def\breakon{\end{multicols}\widetext\vspace{-.2cm}
\noindent\rule{.48\linewidth}{.3mm}\rule{.3mm}{.3cm}\vspace{.0cm}}

\def\breakoff{\vspace{-.2cm}
\noindent
\rule{.52\linewidth}{.0mm}\rule[-.27cm]{.3mm}{.3cm}\rule{.48\linewidth}{.3mm}
\vspace{-.3cm}
\begin{multicols}{2}
\narrowtext}

\draft 

\title{Electron properties of Carbon nanotubes in the field effect regime}

\author{D. S. Novikov and L. S. Levitov}
\address{
\mbox{Physics Department, Center for Materials Sciences \& Engineering,
Massachusetts Institute of Technology, 
Cambridge, MA 02139}
}
 
\maketitle      

\begin{abstract}
Electron properties of Carbon nanotubes can change qualitatively 
in a transverse electric field. 
In metallic tubes the sign of Fermi velocity 
can be reversed in a sufficiently strong field, 
while in semiconducting tubes the  effective mass can change sign.
These changes in the spectrum manifest themselves in a breakup 
of the Fermi surface and in the energy gap suppression, respectively. 
The effect is controlled by the field inside the tube which is screened 
due to the polarization induced on the tube. 
The theory of screening links it with the chiral anomaly for $1D$ fermions 
and obtains a universal screening function determined
solely by the Carbon $\pi$ electron band.
\vskip2mm
\end{abstract}

\bigskip
 
\begin{multicols}{2}
\narrowtext
 

The possibility to change electron spectrum of Carbon nanotubes  
by external field is of interest for basic research as well as 
for nanoscale device engineering. Carbon nanotube (NT) is a $1D$ metal 
or semiconductor depending on the chiral angle\cite{Dresselhaus}. 
Metallic behavior can be suppressed 
by parallel magnetic field that induces a minigap
at the band crossing\cite{B-paral}.
Similar minigaps appear in nominally metallic chiral NTs
due to the intrinsic curvature\cite{KaneMele97,exp-curvature,curv-gaps-exp}.
Novel properties are predicted for the BN tubes
having no inversion symmetry\cite{Mele01,Kral00}.

Here we examine the changes induced in the NT electron spectrum 
by transverse electric field ${\cal E}$ 
strong enough to mix different NT subbands:
\be\label{eq:Erequired}
e{\cal E}R \simeq \Delta_0\equiv \hbar v/R
,\quad
{\cal E}\,[{\rm MV/cm}] \simeq 5.26/R^2\,[{\rm nm}^2],
\ee
where $R$ is the tube radius and $v$ is electron velocity. 
In such a field the effect on electron spectrum is dramatic:
in metallic tubes the electron velocity $v=d\epsilon/dp$
can be reduced and even reverse the
sign, causing Fermi surface breakup, while
in semiconducting tubes the effective mass sign can change, 
which is accompanied by strong suppression of the excitation gap
(Fig.\ref{fig:MNT,SNT}). 

The NT electron system in this regime can be a host of intriguing
many-body phenomena. The reduction of electron velocity in metallic tubes 
leads to an increase of the dimensionless interaction $e^2/\hbar v$ 
that controls the Luttinger liquid properties\cite{all-Luttinger}.
One expects this to enhance the intrinsic, interaction-induced energy gap 
predicted to be small
in pristine NTs\cite{all-Luttinger}. 

Even more peculiar is the 
negative $v$ state with intertwining electron and hole Fermi surfaces 
(Fig.\ref{fig:MNT,SNT} top). This system provides a 
realization of a metallic state unstable with respect
to electron-hole pairing into excitons. Such an instability,
long-envisioned \cite{excitonic} by Mott, Keldysh and Kopaev, and others, 
is especially interesting for the 
mirror symmetric electron and hole 
bands, described by an analog of the BCS theory. 
The chiral gauge symmetry (\ref{eq:chiralGtrans}) of the NT electron
Hamiltonian discussed below
eliminates the interband matrix elements of particle density. 
This makes the phase of the excitonic order parameter 
a gapless Goldstone field, similar to the BCS order parameter. 
Despite its simplicity, no system with such properties has been 
unambiguously identified so far, and it is thus possible that nanotubes 
in a transverse field present a unique opportunity to study this 
phenomenon. 

The field (\ref{eq:Erequired}) required to create this state is to be 
achieved within the tube where the external field is partially 
screened\cite{Benedict95}.
However, because of the discrete bands with relatively large 
interband separation
$\Delta_0\,[{\rm eV}]\equiv \hbar v/R=0.53/R\,[{\rm nm}]$, 
the transverse field penetrates in the NT fairly well. 
We find the screening factor to be $R$-independent and close to $5$
(in accord with \cite{Benedict95}), 
for both metallic and semiconducting tubes.
With the screening taken into account the numbers 
for the required field remain feasible.
(Fields up to $50\,{\rm MV/cm}$ have been 
demonstrated recently in $2D$ gated structures\cite{Battlog}.)

Unexpectedly, the problem of transverse field screening has 
a relation with the chiral anomaly. 
A significant part of electron energy in an external field, in the 
massless Dirac model (standard for NT),
arises from the
effects at the Fermi sea bottom, where a regularization of this
model is required. 
However, the anomaly links the regularized energy
with the properties near the Fermi level (the number of fermion 
species and their velocity),
and thereby generates 
a universal screening function determined only 
by the Carbon $\pi$ electron band. 

We consider the NT at half-filling ignoring charge accumulation
due to gating. Gating in itself 
will not modify the transverse field within the tube, since
a uniformly charged cylinder 
is equipotential. Charging may affect
the inner NT field indirectly via changing screening, 
but this effect should not be significant at moderate gating. 

The $\pi$ electron Carbon band is described by a nearest-neighbor
tight-binding problem 
$\epsilon \psi_r=t\sum_{r'}\psi_{r'}$
with the hopping amplitude $t\simeq 3\,{\rm eV}$.
The states
with small $|\epsilon|\ll t$
are described, 
separately at each of the $K$ and $K'$ points, by a massless Dirac 
Hamiltonian
\be\label{eq:DiracKK'}
{\cal H}_0=-i\hbar v(\sigma_y \partial_x-\sigma_x \partial_y)
,\quad v=3ta_{\rm c-c}/2\hbar
\ee
For a NT in the presence 
of a transverse electric field ${\cal E}$,
\be\label{eq:Hdirac}
{\cal H}_0=\hbar v \lp i\sigma_x \partial_y + \sigma_y k \rp 
-e{\cal E}R \cos (y/R)
\ee
with $k$ the longitudinal momentum and $R$ the NT radius. The 
boundary conditions are quasiperiodic:
\be\label{eq:q-periodicity}
\psi(y+2\pi R)=e^{2\pi i\delta}\psi(y)
\ ,\quad 
\delta=\cases{0\ ,& metallic\cr
\pm\frac13\ ,& semic.}
\ee
The effects of NT curvature\cite{KaneMele97} as well as of 
a parallel magnetic field\cite{B-paral} can in be included 
by slightly shifting $\delta$ away from the ideal values 
(\ref{eq:q-periodicity}). 

We employ a chiral gauge transformation
\be\label{eq:chiralGtrans}
\psi(y)=e^{-i\sigma_x\phi(y)}\widetilde\psi(y)
\ ,\quad
\phi(y)=\frac{e{\cal E}R^2}{\hbar v}\sin (y/R)
\ee
which preserves the condition (\ref{eq:q-periodicity}) and 
turns Eq.(\ref{eq:Hdirac}) into
\be\label{eq:Htransformed}
\widetilde{\cal H}_0=\hbar v \lp i\sigma_x \partial_y + k e^{2i\sigma_x\phi(y)}\sigma_y \rp 
\ee
The transformed Hamiltonian reveals that, 
in particular, the spectrum at $k=0$ is totally independent of 
the transverse field (Fig.\ref{fig:MNT,SNT}).

\begin{figure}[t]
\centerline{\psfig{file=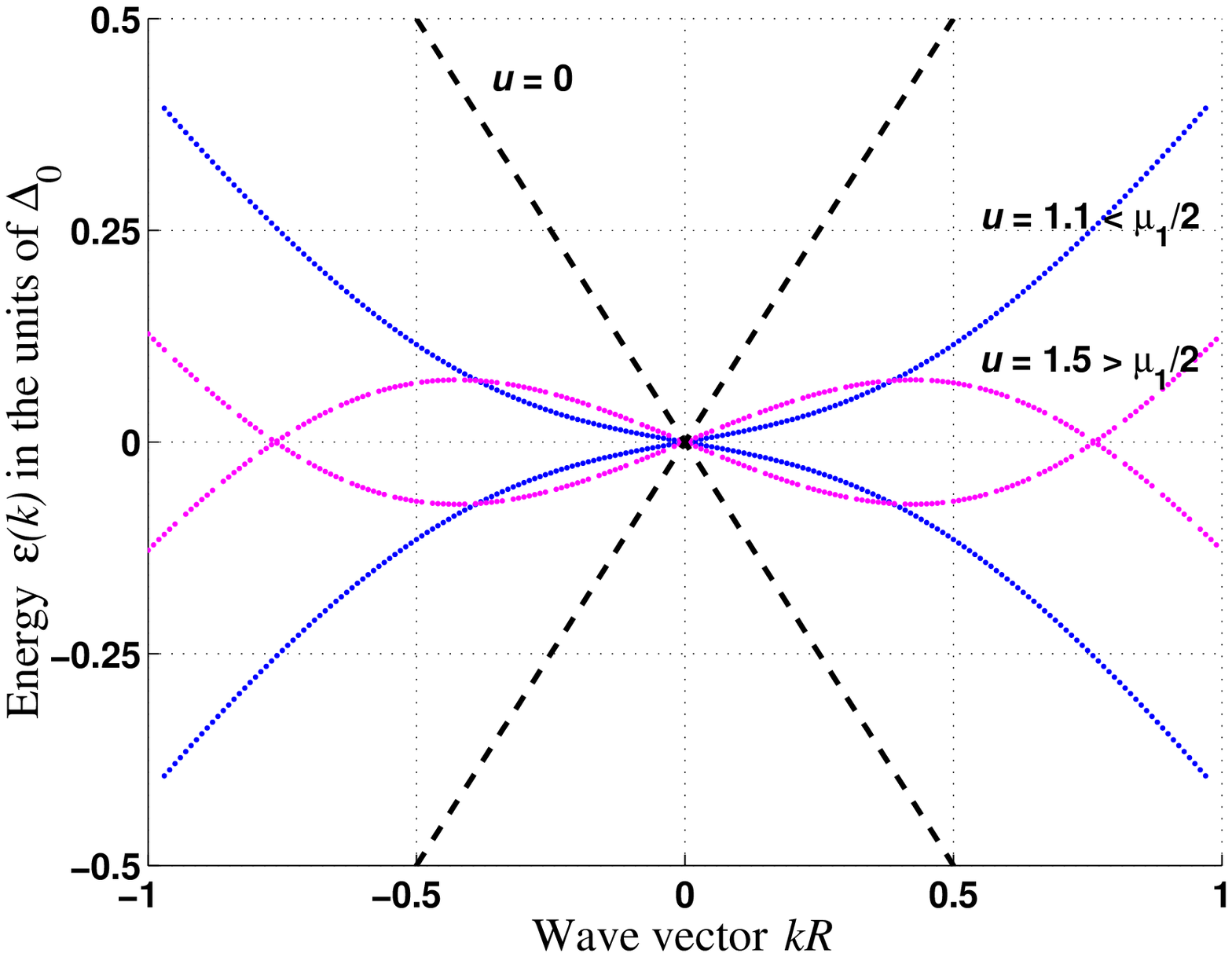,width=3.5in,height=2.5in}}
\vspace{-3mm}
\centerline{
\begin{minipage}[t]{3.5in}
\vspace{0pt}
\centering
\includegraphics[width=3.5in]{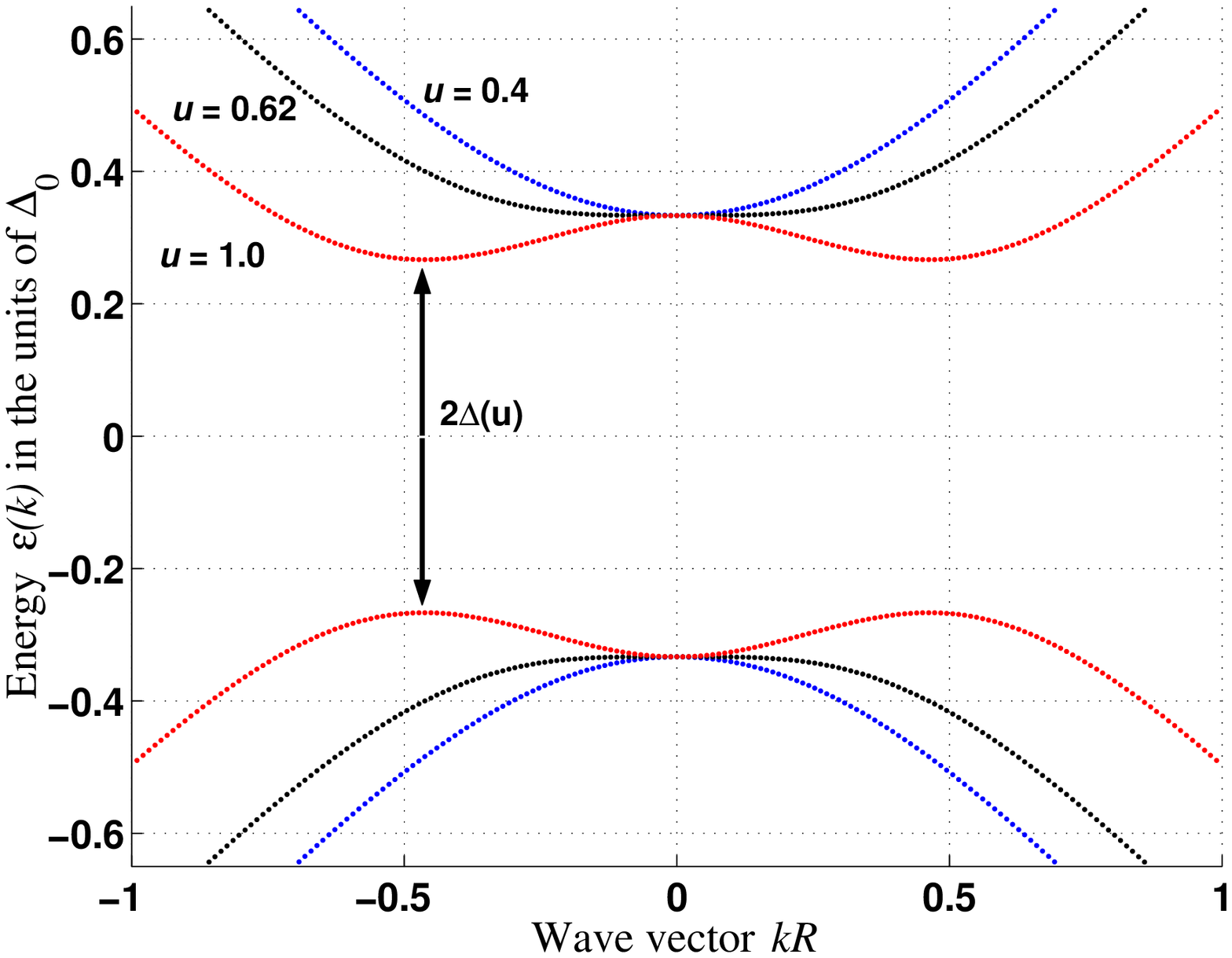}
\end{minipage}
\hspace{-1.5in}
\begin{minipage}[t]{1.4in}
\vspace{0pt}
\centering 
\includegraphics[width=1.4in]{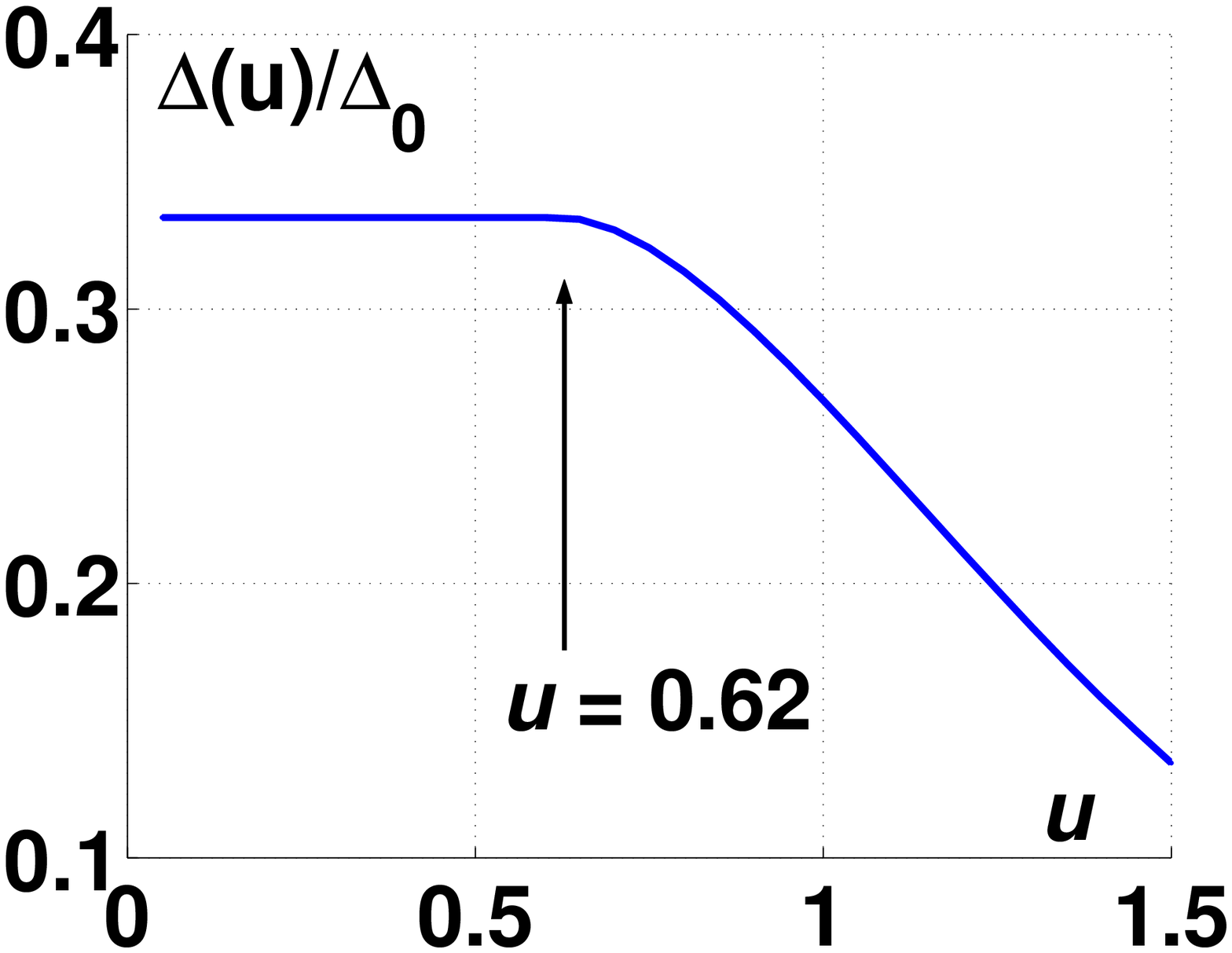}
\end{minipage}
}
\vspace{0.5cm}
\caption[]{
Electron bands transformation:
velocity reversal in metallic NT (top);
effective mass sign change in semiconducting NT (bottom). 
The bands are shown for the dimensionless transverse field 
$u=e{\cal E}R/\Delta_0$ 
below and above critical. 
{\it Inset:} Energy gap suppression in a semiconducting NT.
  }
\label{fig:MNT,SNT}
\end{figure}

The spectrum of a metallic NT
near the band crossing at $kR\ll1$ can be found by employing the 
degenerate perturbation theory. We project the Hamiltonian 
(\ref{eq:Htransformed})
on the two states $|\uparrow\rangle$, 
$|\downarrow\rangle$ degenerate at $k=0$, and obtain
\be\label{eq:J0,u}
\epsilon(k)=\pm \lb \hbar v J_0(2u)\rb\, k
\ ,\quad
u\equiv e{\cal E}R/\Delta_0 = e{\cal E}R^2/\hbar v
\ee
Electron velocity changes sign at the roots of  
the Bessel function $J_0$, first at $2u=\mu_1\approx 2.405$ 
(Fig.\ref{fig:MNT,SNT}). At $u$ above critical the Fermi surface 
fractures: an additional small pocket appears for each spectral 
branch.

The level shifts in semiconducting NT at small $k$ are given by 
the second order perturbation theory in the $k$-term of 
the transformed Hamiltonian (\ref{eq:Htransformed}):
\bea
&& \epsilon_n^\pm(k)=\pm\Delta_0 \lp (n+\delta)+A_n (kR)^2 \rp
\\
&& A_n=
\sum_{m=-\infty}^\infty J_m^2(2u)\frac{2(n+\delta)}{4(n+\delta)^2-m^2}
\eea
For $\delta=1/3$ the curvature of the lowest band $A_0$ changes sign at
$u_c\approx 0.6215$ (Fig.\ref{fig:MNT,SNT}).
This leads to a singular behavior of the excitation gap
which is constant at $u<u_c$ and sharply decreases at $u>u_c$ 
(Fig.\ref{fig:MNT,SNT} inset). This occurs because of 
the lowest excitation energy shifting from $k=0$ at $u<u_c$ 
to $k\ne 0$ at $u>u_c$. The threshold-like suppression of the gap
can be detected by a transport measurement in a thermally activated regime.

The chiral gauge symmetry (\ref{eq:chiralGtrans}) that protects 
the spectrum at $k=0$ is a distinct feature of 
the Dirac model (\ref{eq:DiracKK'}), (\ref{eq:Hdirac}). 
The $\pi$ electron tight-binding 
problem, in the next-lowest gradient order, 
generates a correction to the Hamiltonian (\ref{eq:DiracKK'})
violating the symmetry (\ref{eq:chiralGtrans}):
\be\label{eq:H1}
{\cal H}={\cal H}_0+\lambda
e^{-\frac{3i}2\theta \sigma_z}\!\! \left[ 
 \sigma_x\left(\partial_x^2\!-\!\partial_y^2\right)
\! -\!2\sigma_y\partial_x\partial_y
\right] e^{\frac{3i}2\theta \sigma_z}
\ee
with $\lambda\!=\!\frac14 a_{\rm c-c}\hbar v$ and 
$\theta$ the NT chiral 
angle\cite{Dresselhaus}. 
The transformation (\ref{eq:chiralGtrans}) applied to (\ref{eq:H1})
gives a minigap\cite{E-perp} 
\be\label{eq:gap-perp}
\Delta_{\cal E} = |\sin 3\theta|\, (a_{\rm c-c}/8\hbar v)\,(e{\cal E}R)^2
\ee
Since $8R \gg a_{\rm c-c}$, 
the minigap (\ref{eq:gap-perp})
is too small to alter the behavior at the energies of 
interest, $\epsilon\simeq\Delta_0$.

The main effect of electron interaction is screening of the inner field that 
couples to the electron motion. Here we derive the relation between 
the inner and outer fields.
We first show how the screening problem is reduced to the calculation
of the NT electron energy in the presence of an external field. 
Hereafter we measure all energies in the units of $\Delta_0=\hbar v/R$
and use dimensionless field $u=e{\cal E}R/\Delta_0$.
From Gauss' law, the fields inside and outside the tube are related 
with the induced surface charge density per one fermion species 
(spin and valley) by
\be\label{eq:Gauss}
{\cal E}_{\rm ext}={\cal E}+{\textstyle \frac12} \cdot 4\pi \cdot 4\sigma
\ee
where the factor $1/2$ accounts for depolarization in the cylindrical geometry. 
In Eq.(\ref{eq:Gauss}) we projected the actual charge density on 
the $\cos\varphi$ harmonic as $\sigma(\varphi) \to 4\sigma \cos\varphi$,
ignoring the higher order harmonics. (Here $\varphi\equiv y/R$.)

To obtain the $\cos\varphi$ harmonic of the induced charge, 
we evaluate the dipole moment per unit length 
as $P=- dW({\cal E})/d{\cal E}$, where $W({\cal E})$ is 
the energy of one fermion species
as a function
of the inner field. Combining this with the relation
$\sigma=P/(\pi R^2)$
and with the Gauss' law (\ref{eq:Gauss}),
and passing to dimensionless $u_{\rm ext}$, $u$,
we obtain 
\be\label{eq:UU'W}
u_{\rm ext}=u+8\frac{e^2}{\hbar v}P(u)
\ee
After the dipole moment $P(u)$ is known
Eq.(\ref{eq:UU'W}) can be solved for the inner field $u$ in terms 
of the outer field $u_{\rm ext}$.

We consider the general problem of electron energy 
in a transverse field in a free particle model.
The electron levels $\epsilon_{n,k}$ perturbed by the field
can be easily found numerically 
at each value of the longitudinal momentum
$k$
by using a transfer matrix for Eq.(\ref{eq:Htransformed}).
The level shifts 
$\delta \epsilon_{n,k}=\epsilon_{n,k}(u)-\epsilon_{n,k}^{(0)}$ decrease at large $|n|$, and 
the series 
the total change of the occupied states energy 
\be\label{E0}
E_0(k)=\sum\limits_{n=-\infty}^{+\infty} \delta \epsilon_{n,k}
\quad ({\rm such\ that\ } \epsilon_{n,k}(u)<0)
\ee
rapidly converge at $n\to\pm\infty$. 

There are two basic problems with Eq.(\ref{E0}):
1) Due to an upward shift of the filled levels (Fig.1), 
$E_0$ is positive and also has positive derivative $dE_0/du$.
Hence Eq.(\ref{E0}) leads to the dipole moment $P_0=-dE_0/du$ 
opposite to the field, i.e. to an
unphysical ``diamagnetic'' polarization sign instead of 
the expected ``paramagnetic'' effect. 
2) The dependence of the energy $E_0$ on the longitudinal wavevector $k$
leads to an ultraviolet divergence in the integral
$P=\int P(k)dk$, because $E_0(k)$ increases with $|k|$, saturating
at $|k|R\gg1$ at an asymptotic value $\frac12 u^2$.

Both difficulties are resolved by taking into account
a fundamentally important contribution to the energy
that arises due to the effects at the {\it Fermi sea bottom}. 
Physically, the finite electron band width 
invalidates the massless Dirac 
approximation at large negative energies. 
This contribution, however,
depends solely on the number of Dirac fermion species and their velocity $v$,
and is totally insensitive to any other details including 
the longitudinal momentum $k$ value. We find that
\be\label{eq:Eanomaly}
E_{\rm anom}=-{\textstyle\frac12}u^2
\ee
for each fermion species.
Remarkably, Eq.(\ref{eq:Eanomaly}) can be obtained without detailed discussion 
of the behavior at the interatomic length scales 
--- the universality of 
Eq.(\ref{eq:Eanomaly}) is rooted in the physics of {\it the chiral anomaly}
in the $1+1$ fermion problem.
The resulting total energy integral
\be\label{eq:Wintegral}
W = \int_{-\infty}^\infty \lp E_0(k)+E_{\rm anom}\rp {\textstyle \frac{dk}{2\pi}}
\ee
converges after $E_0(k)$ is offset by $E_{\rm anom}$ (Fig.\ref{fig:P(k)}).

We evaluate the energy (\ref{E0}) and derive the anomaly (\ref{eq:Eanomaly})
for a weak field $u\ll 1$. The NT bands at $u=0$ are
\be
\epsilon_n^\pm(k)=\pm\sqrt{(n+\delta)^2+k^2}\ ,\quad -\infty<n<+\infty
\ee
In a half-filled system with just the $\epsilon_n^-(k)\!\!<\!\!0$ bands
filled, the external field $u$ changes the
Fermi sea energy by
\be
W=- \int {\sum_n}'
\delta \epsilon_n^-(k) \frac{dk}{2\pi}
\ee
Here the superscript in $\sum'$ indicates 
regularization by truncating the interaction with 
the external field at a certain large negative  energy. 
We check that this contribution 
to the energy is independent of 
the details of truncation and obtain the anomaly (\ref{eq:Eanomaly}) 
by choosing a convenient truncation scheme.

\begin{figure}
\centerline{\psfig{file=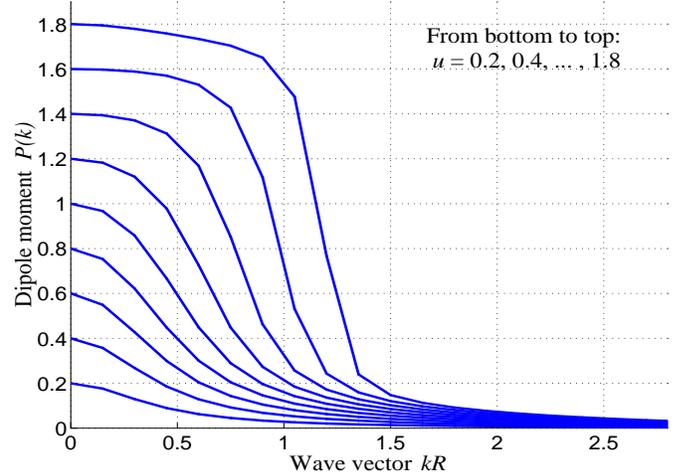,width=3.5in,height=2.5in}}
\vspace{0.1cm}
 \caption[]{
Dipole moment $P(k)=-d(E_0(k)+E_{\rm anom})/du$ 
per one fermion species in a semiconducting NT 
as a function of $k$. Note that the energy anomaly (\ref{eq:Eanomaly})
cancels with $E_0(k)$ 
at $kR\gg1$, assuring convergence of $P_{\rm total}=\int P(k)dk/2\pi$. 
Note also that $P(k\to 0)$ is dominated by the anomaly, since
$E_0=0$ at $k=0$ due to the chiral gauge invariance
(\ref{eq:chiralGtrans}).
  }
\label{fig:P(k)}
\end{figure}

The level shifts $\delta \epsilon_n^-(k)$, in the second order of 
the perturbation theory in the external field 
$\hat V\!\!=\!\!-e{\cal E}R\cos\varphi$, are
\be\label{eq:sum-perturb}
\delta \epsilon_n^-=
\sum_{m}\frac{|\langle m^+|\hat V|n^-\rangle|^2}{\epsilon_n^- -\epsilon_m^+}
+ \sum_{m} 
\frac{|\langle m^-|\hat V|n^-\rangle|^2}{\epsilon_n^- -\epsilon_m^-}
\ee
where the superscript $\pm$ indicates the electron and hole branches 
and the $k$ dependence is suppressed. Due to the integration over $\varphi$
with $\hat V\propto\cos\varphi$ in the matrix elements the only nonzero 
terms in (\ref{eq:sum-perturb}) are those with $m=n\pm 1$. 

We now show that the sums over $\epsilon_m^+$ and $\epsilon_m^-$ in 
(\ref{eq:sum-perturb}), respectively, give the regular and the anomalous 
contributions to the total energy $W=\sum_n'\delta \epsilon_n^-$. 
Different behavior of the two sums under regularization stems
from their different convergence type. Individual terms in the sum over
$\epsilon_m^+$ decrease rapidly at large $m$, so that 
the series for $W$
is absolutely convergent. On the other hand, in the sum 
over $\epsilon_m^-$ the terms do not change at large $m$ and thus the 
corresponding contribution to $W$ is given by poorly
convergent and regularization-sensitive series.

Taking from (\ref{eq:sum-perturb}) just the terms with $\epsilon_m^+$,
evaluating the matrix elements $\langle m^+|\hat V|n^-\rangle$ and
summing over $n$ yields
\be\label{eq:Ereg-perurb}
E_0(k) = 
\frac{u^2}4 \sum_n \frac{\epsilon_n^+(k)\epsilon_{n'}^+(k)\!-\!(n\!+\!\delta)(n'\!+\!\delta)\!-\!k^2}{\epsilon_n^+(k)\epsilon_{n'}^+(k)(\epsilon_n^-(k) - \epsilon_{n'}^+(k))}
\ee
with $n'=n+1$. The sum (\ref{eq:Ereg-perurb}) rapidly converges at large $n\to\pm\infty$
and can be easily evaluated numerically.

Now we consider the sum of the level shifts $W={\sum}_n'\delta \epsilon_n^-$ 
taking into account only the second term in
(\ref{eq:sum-perturb}). At the first sight this sum is identically zero. 
Indeed, due to the symmetry 
$\langle m^-|\hat V|n^-\rangle = \langle n^-|\hat V|m^-\rangle$,
in the sum over $n$ with $m=n\pm1$ all the terms cancel in pairs. 
However, truncation of the interaction at a large negative energy
compromises the cancellation and yields a finite result. If one sets
$\langle m^-|\hat V|n^-\rangle =0$ for all 
$|m|$ or $|n|$ exceeding a large number $N$, there will be just two terms
in the sum over $n$ that do not cancel:
\be\label{eq:Nterms}
 E_{\rm anom}=
\frac{|\langle N'^-|\hat V|N^-\rangle|^2}{\epsilon_N^- -\epsilon_{N'}^-}
+
\frac{|\langle \!-\!N'^-|\hat V|-\!N^-\rangle|^2}{\epsilon_{-N}^- -\epsilon_{-N'}^-}
\ee
with $N'=N-1$.
Evaluating the matrix elements and energy levels is straightforward because
at large $N\gg |k|$ one can set $k=0$.
The result, coinciding with (\ref{eq:Eanomaly}), is robust
under a change of the regularization.

The expression (\ref{eq:Eanomaly}) for the energy anomaly, 
derived above for the weak field, is in fact more general. To illustrate 
this we consider a special case of zero longitudinal momentum $k=0$ 
and derive (\ref{eq:Eanomaly}) from bosonization, without using 
perturbation theory in $u\ll 1$. After the problem (\ref{eq:Hdirac}) 
is bosonized in the standard way\cite{Stone}, 
using $\psi_{L,R}\propto  e^{i\phi_{L}},  e^{-i\phi_{R}}$,
we obtain  
a quadratic Hamiltonian
  \be\label{H-bos}
{\cal H} =  \int_0^{2\pi R} 
\sum\limits_{j=L,R}\lb
{\textstyle \frac{\hbar v}{4\pi}} (\partial_{x}\phi_{j})^2 
+{\textstyle \frac{e}{2\pi}}\partial_{x}\phi_{j}U(y)\rb d y 
  \ee
The second term in (\ref{H-bos}) representing interaction 
with the external field $U(y)$ can be decoupled by a shift
$\phi_j \to \phi'_j - \frac1{\hbar v_F}\int_0^y U(y')dy'$.
The Hamiltonian for $\phi'_j$ takes the 
form (\ref{H-bos}) with $U=0$, while the ground state energy
\be
\delta E = - \frac{e^2}{2\pi \hbar v_F} \int_0^{2\pi R}\!\!\! U^2(y)dy
\ee
is nothing but the anomaly (\ref{eq:Eanomaly}) scaled by $\Delta_0=\hbar v/R$.

After adding the energies (\ref{eq:Ereg-perurb}) and (\ref{eq:Eanomaly}), 
and integrating in (\ref{eq:Wintegral}) over $k$ numerically, one obtains
\be\label{Wlinearized}
W=-\frac{\alpha}2u^2,\quad 
\alpha=\cases{0.196... & \ for $\delta=1/3$\cr
0.179... & \ for $\delta=0$}
\ee
Eq.(\ref{eq:Gauss}) with the dipole moment $P = - dW/d{\cal E}$ and
the charge density $\sigma = P/\pi R^2$ yield the screening function
\be\label{u-ratio}
{\cal E}_{\rm ext} = \lp 1+8\alpha \, {\textstyle \frac{e^2}{\hbar v}} \rp {\cal E}
\ee
With $e^2/\hbar v=2.7$ 
this gives ${\cal E}_{\rm ext}/{\cal E} = 5.24$ for $\delta=1/3$, and 
${\cal E}_{\rm ext}/{\cal E} = 4.87$ for $\delta=0$. 
The outer-to-inner field ratio $\simeq 5$ (see \cite{Benedict95} for another 
derivation) renders
the required fields (\ref{eq:Erequired}) feasible\cite{tobepublished}. 
Interestingly, the screening (\ref{u-ratio}) 
is independent of the tube radius $R$ and
is almost the same in the metallic and semiconducting NTs. 
The latter is not surprising, since the screening 
is absent in a single 1D mode approximation: the polarizability 
is related with dipolar transitions between {\it different} subbands. 

The radius-independence of (\ref{u-ratio}) resembles an effect 
of a dielectric constant. We note, however, that the change of the inner field
due to individual Carbon atoms polarizability is small
in $a_{\rm c-c}/2\pi R \ll 1$. The result (\ref{u-ratio})
reflects the semimetallic character of the $\pi$ electron 
band with the density of states vanishing at the band center. 

In summary, nanotube electron states undergo interesting transformations
in the field effect regime, leading to novel phenomena in both 
the single particle and many-body properties. The analysis of 
screening, performed using a relation with the theory of chiral anomaly, 
indicates that the fields required for the observation 
of the proposed effects are in the experimentally feasible range. 

This work was supported by the MRSEC Program
of the National Science Foundation under Grant No. DMR 98-08941.
 

\end{multicols}

\begin{thebibliography}{99} 

\bibitem{Dresselhaus} R. Saito, G. Dresselhaus, and M. S. Dresselhaus, 
{\it Physical Properties of Carbon Nanotubes} 
(Imperial College Press, London, 1998).

\bibitem{B-paral}
H. Ajiki and T. Ando, 
J. Phys. Soc. Jpn. {\bf 65}, 505 (1996);
J. -O. Lee,
J. R. Kim, J. J. Kim, J. Kim, N. Kim, J. W. Park, and K. H. Yoo,
Sol. Stat. Comm. {\bf 115}, 467 (2000).

\bibitem{KaneMele97} 
C. L. Kane, E. J. Mele,
Phys. Rev. Lett. {\bf 78}, 1932 (1997).

\bibitem{exp-curvature}
C. Zhou, J. Kong, H. Dai,
Phys. Rev. Lett. {\bf 84}, 5604 (2000).

\bibitem{curv-gaps-exp}
M. Ouyang, J. L. Huang, C. L. Cheung, C.M. Lieber,
Science {\bf 292}, 702 (2001)

\bibitem{Mele01}
E. J. Mele, P. Kral,
Phys. Rev. Lett. 88, 056803 (2002);

\bibitem{Kral00}
P. Kral, E. J. Mele, D. Tomanek,
Phys. Rev. Lett. 85, 1512 (2000)

\bibitem{all-Luttinger} 
C. Kane, L. Balents, M. P. A. Fisher,
Phys. Rev. Lett. 79, 5086 (1997);
R. Egger, A. O. Gogolin, Phys. Rev. Lett. 79, 5082 (1997);
L. Balents, M. P. A. Fisher, Phys. Rev. B 55, R11 973 (1997);
Yu. A. Krotov, D.-H. Lee, S.~G.~Louie, Phys. Rev. Lett. 78, 4245 (1997).

\bibitem{excitonic} 
N. F. Mott, Phil. Mag. 6, 287 (1961); 
L. V. Keldysh, Yu. V. Kopaev, Fiz. Tverd. Tela 6, 2791 (1964); [Soviet Phys.-Solid State 6, 2219 (1965)];
J. J. des Cloizeaux, Phys. Chem. Solids 26, 259-266 (1965);
B. I. Halperin, T. M. Rice, in Solid State Physics 21, 115-192 (1968).

\bibitem{Benedict95}
L. X. Benedict, S. G. Louie, and M. L. Cohen,
Phys. Rev. B{\bf 52}, 8541 (1995)

\bibitem{Battlog} 
J. H. Sch\"on, S. Berg, Ch. Kloc, B. Battlog,
Science {\bf 287}, 1022 (2000);
J. H. Sch\"on, A. Dodabalapur, Z. Bao, Ch. Kloc, O. Schenker, B. Battlog,
Nature {\bf 410}, 189 (2001)

\bibitem{Stone} M. Stone (ed.), ``Bosonization'' (World Scientific, 1994)

\bibitem{tobepublished}
For the field $u\simeq 1$ the screening factor
is close to that found in the weak field $u\ll1$ (to be published).





\bibitem{E-perp}
The minigap (\ref{eq:gap-perp}) is $\simeq 8R/a_{\rm c-c}$ times 
smaller than that found in
X. Zhou,
Hu Chen, Ou-Yang Zhong-can,
J. Phys.-Cond. Mat. {\bf 13} (27), L635 (2001)



\end{thebibliography}
\end{document}